Title: Quantum Finite State Automata

Authors: Debayan Ganguly[1], Kumar Sankar Ray[2]

Affiliations: [1,2]Electronics and Communication Sciences Unit, Indian Statistical Institute, Kolkata-108.

# 2-tape 1-way Quantum Finite State Automata

*Debayan Ganguly[1], Kumar Sankar Ray[2]*

[1,2]*Electronics and Communication Sciences Unit, Indian Statistical Institute, Kolkata-108.*

[1]debayan3737@gmail.com , [2]ksray@isical.ac.in

*Abstract: 1-way quantum finite state automata are reversible in nature, which greatly reduces its accepting property. In fact, the set of languages accepted by 1-way quantum finite automata is a proper subset of regular languages. We introduce 2-tape 1-way quantum finite state automaton (2T1QFA(2))which is a modified version of 1-way 2-head quantum finite state automaton(1QFA(2)). In this paper, we replace the single tape of 1-way 2-head quantum finite state automaton with two tapes. The content of the second tape is determined using a relation defined on input alphabet. The main claims of this paper are as follows: (1)We establish that 2-tape 1-way quantum finite state automaton(2T1QFA(2)) can accept all regular languages (2)A language which cannot be accepted by any multi-head deterministic finite automaton can be accepted by 2-tape 1-way quantum finite state automaton(2T1QFA(2)) .(3) Exploiting the superposition property of quantum automata we show that 2-tape 1-way quantum finite state automaton(2T1QFA(2)) can accept the language L={ww |w∈ $\{a, b\}^*$}.*

## 1. INTRODUCTION

The number of transistors in a dense integrated circuit doubles approximately every eighteen months known as Moore's law [1]. It is predicted that quantum computer, will replace existing classical model. We know that finite state automaton is one of the simplest models of computation for classical computer. Similarly, quantum finite state automaton is the basic model of computation for quantum computers.

Since measurement is an important operation in quantum computation and quantum information, two main methods of measurement have been investigated in quantum automata. Measure once quantum finite state automaton introduced by Moore and Crutchfield [2] where a measurement is performed at the end of computation and measure-many quantum finite state automaton introduced by Kondacs and Watrous [3] where a measurement is performed after each discrete step. Measure-many quantum finite state automaton can accept all languages that can be accepted by measure-once quantum finite state automaton [4]. So, in this paper, whenever we mention 1-way quantum finite state automaton (1QFA), it is the model introduced by Kondacs and Watrous [3]. The basic models of classical finite state automaton accept all regular languages but set of languages accepted by 1QFAs with bounded error is a proper subset of regular languages [3]. In search of increasing the language accepting capabilities, different models of 1QFA has been proposed [5-11] by researchers.

Multihead finite state automata and their language accepting capabilities are shown in [12-13]. Kutrib et.al. [14] explored the computational power of one-way multihead reversible and proved that one-way multihead finite automaton with two heads can accept all uniletter regular languages. Two-way reversible multihead finite automata and its language accepting capabilities are studied by Morita [15].

Ambainis et.al. [16] Introduced the notion of quantum finite multitape automata and proved that there is a language recognized by a quantum finite multitape automaton but not by deterministic or probabilistic finite automaton. In [17], they showed that multiletter-QFA accept a language not recognizable by a 1-way quantum finite state automata(1QFA).The decidability of the equivalence and minimization problems of multiletter QFAs has been studied in [18]. Further, Qiu et.al [19] showed that (k+1)-letter QFAs are computationally more powerful than k-letter QFA and they studied the equivalence of multiletter QFA. Zheng,S et.al [20] introduced 2-tape finite automata with quantum and classical states and proved that $\{a^n b^{n^k} | n \in N\}$ can be recognized by this model. Ganguly, D et.al. [21] proposed a model,namely,1-way multihead quantum finite state automaton(1QFA(k)) by introducing multiple heads combined with existing automaton and studied its language recognizing capabilities. It is proved that the model 1QFA(2) can accept all unary regular languages and 1QFA(2) is more powerful than 1-way reversible 2-head finite state automaton(1RMFA(2)).

In this paper, we introduce a new model of one-way quantum finite state automaton, namely 2-tape 1-way quantum finite state automaton (2T1QFA(2)) which is a one-way quantum finite automaton where the single input tape of 1QFA(2) is replaced by two tapes with two independent heads. Quantum finite automata are inherently reversible in nature and it is this absence of non-determinism in the model which greatly reduces their computational power. The content of the second tape is determined by a relation defined on input alphabet. The two tapes and the relation defined on input alphabet in the 2-tape 1-way quantum finite state automaton (2T1QFA(2)) enable us to introduce non-determinism in the quantum model. We primarily shift the non-determinism from the automaton to the input. Thus, the automaton still remains reversible and retains its quantum properties. We show that 2-tape 1-way quantum finite state automaton (2T1QFA(2)) in spite of being reversible accept all regular languages. We further show that the above mentioned model is more powerful than one-way reversible 2-head finite automata. In fact, 2-tape 1-way quantum finite state automaton (2T1QFA(2)) accept language which is not accepted by any multi-head deterministic finite automaton. Moreover by exploiting the superposition property, we show that 2-tape 1-way quantum finite state automaton (2T1QFA(2)) can accept the language L={ww |w∈ {a, b}$^*$}.

## 2. PRELIMINARIES

In this section, we give different definitions and corresponding results for 1QFA.

### 2.1 Quantum finite state automata

One-way quantum finite state automaton can been seen as the simplest model of quantum computation.

#### 2.1.1  1-way quantum finite state automata

1QFA are very simple but less powerful than classical 1-way finite automaton.

**Measure many quantum finite state automata(1QFA)**

We consider 1-way quantum finite automata (QFA) as defined in [2].

**Definition 1** *Namely, a 1-way QFA is a six tuple $M = (Q, \Sigma, \delta, q_0, Q_a, Q_r)$ where*

*a. Q is a finite set of states,*

*b. Σ is an input alphabet,*

*c. δ is a transition function,*

*d. $q_0 \in Q$ is a starting state*

*e. $Q_a \in Q$ and $Q_r \in Q$ are sets of accepting and rejecting states.*

*The states in $Q_a$ and $Q_r$ are called halting states and the states in $Q_{non} = Q - (Q_a \cup Q_r)$ are called non-halting states. The symbols # and $ do not belong to Σ. We use # and $ as the left and the right end marker in both the tapes, respectively. The working alphabet of M is Γ=Σ ∪{#,$}.*

A superposition of M is any element of l$_2$(Q)(the space of mappings from Q to ℂ with l$_2$ norm). For q∈Q, |q⟩ denotes the unit vector with value 1 at q and 0 elsewhere. All elements of l$_2$(Q) can be expressed as a linear combination of vectors |q⟩. We will use ψ to denote l$_2$(Q).

The transition function δ maps Q×Γ×Q to ℂ where ℂ denotes the set of complex numbers. The value $\delta(q_1, a, q_2)$ is the amplitude of |q$_2$⟩ in the superposition of states to which M goes from |q$_1$⟩ after reading '$a$'. For $a \in \Gamma$, $V_a$ is a linear transformation on l$_2$(Q) defined by

$$V_a(|q_1\rangle) = \sum_{q_2 \in Q} \delta(q_1, a, q_2)|q_2\rangle.$$

We require all $V_a$ to be unitary.

The computation of a one-way quantum finite automaton starts in the superposition |q$_0$⟩. Then transformations corresponding to left endmarker '#', the letters of the input word w and the right endmarker '$' are applied.
The transformation corresponding to $a \in \Gamma$ consists of two steps.
1)First, $V_a$ is applied. The new superposition ψ' is $V_a(\psi)$ where ψ is the superposition before this step.
2) Then, ψ' is observed with respect to the observable E$_{acc}$⊕E$_{rej}$⊕E$_{non}$ where E$_{acc}$=span {|q⟩:q∈Q$_{acc}$}, E$_{rej}$=span{|q⟩:q∈Q$_{rej}$}, E$_{non}$=span{|q⟩:q∈Q$_{non}$}. This observation gives x∈E$_i$ with probability equal to the amplitude of the projection of ψ'. After that the superposition collapses to the projection.
If we get ψ'∈ E$_{acc}$, the input is accepted. If ψ'∈ E$_{rej}$, the input is rejected. If ψ'∈ E$_{non}$, the next transformation is applied.
We regard these transformations (1) & (2) as reading a letter '$a$'.
The above stated definition of 1QFA is from [2].

#### 2.1.2   1-way 2-head quantum finite state automata(1QFA(2))

**Definition 2.** *A 1-way 2-head quantum finite state automaton is a automaton $M = (Q, Q_a, |q_0>, \Sigma, \delta)$ where*
a. *Q is a finite set of states,*
b. *$Q_a \in Q$ are set of accepting states.*
c. *$|q_0>$ is the initial quantum state superposition obeying normalization condition.*
d. *Σ is an input alphabet.*
e. *δ is a transition function that assign a unitary trasition matrix $V_\Gamma$ on $C^{|Q|}$ to each string $\Gamma \in (\Sigma \cup \{\#, \$\})^2$ where $C^n$ denotes Euclidean space consisting of all n-dimensional complex vectors. So δ is a mapping of the form $\delta: Q \times \Gamma^2 \times \{0,1\}^2 \to C^Q$ is the partial transition function where 1 means to move the head one square to the right and 0 means to keep the head at current square. We use # and $ as the left and the right end marker, respectively.*

A superposition of M is any element in the Hilbert space $l_2(Q)$. For $q \in Q, |q>$ denotes the unit vector with value 1 at q and 0 elsewhere. All elements of $l_2(Q)$ can be expressed as a linear combination of vectors.

The transition function $\delta$ maps $Q \times \Gamma^2 \times \{0,1\}^2$ to $\mathbb{C}^Q$ where $\mathbb{C}^Q$ denotes the set of complex numbers. The value $\delta(q, \sigma_1, \sigma_2, q', d_1, d_2)$ is the amplitude of $|q'>$ in the superposition of states to which M goes from $|q>$ after reading $\sigma_1$ by 1st head, $\sigma_2$ by 2nd head and moving the heads according to $d_1, \& d_2$ respectively. The head movement 0 denotes it stays in its position and 1 denotes head is moved to the right. For $\sigma_1, \sigma_2 \in \Gamma, V_{\sigma_1,\sigma_2}$ is a linear transformation on $l_2(Q)$ defined by

$$V_{\sigma_1,\sigma_2}(|q>) = \sum_{q' \in Q} \delta(q, \sigma_1, \sigma_2, q', d_1, d_2) |q'>$$

We require all $V_{\sigma_1,\sigma_2}$ to be unitary. The check for wellformedness can be done in a similar manner as in [2] in the following way:

Consider the Hilbert space $l_2(Q)$, where Q is the set of internal states of a 1QFA(k) M. Suppose that we have a linear operator $V_{\sigma_1,\sigma_2}: l_2(Q) \rightarrow l_2(Q)$ for each $\sigma_i \in \Gamma, i = 1 \leq i \leq 2$ and a function $D: Q \rightarrow \{1,0\}^2$. Define transition function $\delta$ as:

$$\delta(q, \sigma, q', d_1, d_2) = <q'|V_{\sigma_1,\sigma_2}|q> \text{ when } D(q_i) = d_1, d_2 \quad \ldots\ldots\ldots\ldots\ldots\ldots (2)$$

and

$$\delta(q, \sigma, q', d_1, d_2) = 0 \text{ when } D(q_i) \neq d_1, d_2 \quad \ldots\ldots\ldots\ldots\ldots\ldots (3)$$

Here $<q'|V_{\sigma_1,\sigma_2}|q>$ denotes the coefficient of $|q'>$ in $V_{\sigma_1,\sigma_2}|q>$. Eventually, M is well-formed if and only if

$$\sum_{q'} <q''|V_{\sigma_1,\sigma_2}|q><q''|V_{\sigma_1,\sigma_2}|q'> = 1 \text{ if } q = q'$$

and

$$\sum_{q'} <q''|V_{\sigma_1,\sigma_2}|q><q''|V_{\sigma_1,\sigma_2}|q'> = 0 \text{ if } q \neq q'$$

for each $V_{\sigma_1,\sigma_2}$ pair. pair. The condition mentioned is similar to the condition for reversibility in [15].

The input word w begin with # and ends with $. The input is accepted if and only if the computation halts in an accepting states. It halts when the transition function is not defined for the current situation. In all other cases the input is rejected.

### 3. 2-TAPE 1-WAY QUANTUM FINITE STATE AUTOMATA

A 2-tape 1-way quantum finite state automaton is a quantum finite state automaton with two read only input tapes where content of the second tape is determined by a relation defined on input alphabet whose inscription is the input word in between two end markers. Each tape contains one head which can move to the right or stay on current tape square but not beyond the end markers.

**Definition 3.** *A 2-tape 1-way quantum finite state automaton is a nine tuple $M = (Q, Q_{acc}, Q_{rej}, |q_0>, \Sigma, \delta, \#, \$, \rho)$ where*

a. *Q is a finite set of states,*
b. *$Q_{acc} \in Q$ are set of accepting states.*
c. *$Q_{rej} \in Q$ are set of rejecting states.*
d. *$|q_0>$ is the initial quantum state superposition obeying normalization condition.*
e. *$\Sigma$ is a finite set of input symbols: the tape symbol set is $\Gamma \in \Sigma \cup \{\#, \$\}$; two tapes with two different inputs $w_1$ and $w_2$ are $T_1 = \#w_1\$$ and $T_2 = \#w_2\$$.*
f. *$\delta$ is a transition function that assign a unitary trasition matrix $V_\Gamma$ on $C^{|Q|}$ to each string $\Gamma \in \Sigma \cup \{\#, \$\}$) where $C^n$ denotes Euclidean space consisting of all n-dimensional complex vectors. So $\delta$ is a mapping of the form $\delta: Q \times \Gamma^2 \times \{0,1\}^2 \rightarrow C^Q$ is the partial transition function where 1 means to move the head one square to the right and 0 means to keep the head at current square .*
g. *$\# \notin \Sigma$ is the left and $\$ \notin \Sigma$ is the right endmarkers.*
h. *$\rho$ is a relation defined on input alphabet by which the symbols of the second tape is determined and it depends on the symbols in the first tape. The symbol $\begin{bmatrix}\Sigma\\\Sigma\end{bmatrix}_\rho = \{\begin{bmatrix}a\\b\end{bmatrix} / a, b \in \Sigma, (a, b) \in \rho\}$ and $2T1QFA(2)_\rho(\Sigma) = \begin{bmatrix}\Sigma\\\Sigma\end{bmatrix}_\rho^*$ denotes the 2-tape 1-way quantum finite state automaton associated with $\Sigma$ and $\rho$.*

The states in $Q_{acc}$ and $Q_{rej}$ are called halting states and the states in $Q_{non} = Q - (Q_{acc} \cup Q_{rej})$ are called the non-halting states. The automaton has two input tapes with two heads each on one of the tape where the letters in the corresponding positions on the input tapes are according to the relation $\rho$. The word on the first tape is accepted or rejected by the automaton.

A superposition of M is any element of $l_2(Q)$. For $q \in Q$, $|q)$ denotes the unit vector with value 1 at q and 0 elsewhere. All elements of $l_2(Q)$ can be expressed as a linear combination of vectors $|q)$. We will use $\psi$ to denote $l_2(Q)$.

The transition function δ maps $Q \times \Gamma^2 \times Q \times \{0,1\}^2$ to C where C denotes the set of complex numbers. The value $\delta(q_1, a, b, q_2, d_1, d_2)$ is the amplitude of $|q_2\rangle$ in the superposition of states to which M goes from $|q_1\rangle$ after reading '$a$' in the first tape and 'b' in the second tape and moving the first tape head according to $d_1$ and second tape head according to $d_2$ where zero denotes head stays in its position and one denotes head has moved to the right. For $a, b \in \Gamma$, $V_{a,b}$ is a linear transformation on $l_2(Q)$ defined by $V_{a,b}(|q_1\rangle) = \sum_{q_2 \in Q} \delta(q_1, a, b, q_2, d_1, d_2)|q_2\rangle$. We require all $V_{a,b}$ to be unitary. The check for well-formedness can be done in a similar manner as in [2] in the following way:

Consider the Hilbert space $l_2(Q)$, where Q is the set of internal states of the automaton M. A linear operator $V_{\sigma,\tau}: l_2(Q) \to l_2(Q)$ for each σ,τ pair and a function $D: Q \to \{0,1\}^2$ exist. The transition function δ is defined as

$$\delta(q, \sigma, \tau, q', d_1, d_2) = \begin{cases} \langle q'|V_{\sigma,\tau}|q\rangle & D(q') = (d_1, d_2) \\ 0 & D(q') \neq (d_1, d_2) \end{cases}$$

where $\langle q'|V_{\sigma,\tau}|q\rangle$ denotes the coefficient of $|q'\rangle$ in $V_{\sigma,\tau}|q\rangle$. M is well-formed if and only if $\sum_q \overline{\langle q'|V_{\sigma,\tau}|q_1\rangle} \langle q'|V_{\sigma,\tau}|q_2\rangle = \begin{cases} 1 & q_1 = q_2 \\ 0 & q_1 \neq q_2 \end{cases}$ for each σ, τ pair. The condition mentioned is similar to the condition for reversibility in [13].

The string #$w_1$$ is placed in the first tape and #$w_2$$ in the second tape where $\begin{bmatrix} w_1 \\ w_2 \end{bmatrix} \in 2T1QFA(2)_\rho(\Sigma)$ and the automaton accepts or rejects $w_1$ with some probability. Both tapes begin with # and ends with $.

Note that many values of $V_{\sigma,\tau}|q\rangle$ define transitions which we do not encounter during a computation of w for a particular M. We define those values arbitrarily in such a way that $V_{\sigma,\tau}$ is unitary. In general, we specify only those values that matter for all other values the automaton M goes to some state q where q∈Q, the other values are so assigned that the resulting operator is unitary. So for a state q if no value is mentioned for a pair σ,τ where σ,τ ∈Γ, as Γ is finite therefore number of such σ,τ pairs are also finite. $V_{\sigma,\tau}|q\rangle = |q_{rejq}\rangle$ where defining unmentioned transitions in this way ensures well-formed transitions. Moreover for a given automaton M if such a transition is employed it always rejects. This enables us to define the automaton without mentioning all the σ,τ pairs. We assume these $q_{rejq}$s' belongs to the set $Q_{rej}$ and $D(q_{rejq}) = (0,0)$. As these transitions are included in automaton by default when mentioning the set $Q_{rej}$ and Q we do not explicitly mention these $q_{rejq}$s'.

**Example 1**: $M = (Q, Q_{acc}, Q_{rej}, |q_0\rangle, \Sigma, \delta, \rho)$ is a 2-tape 1-way quantum finite state automaton that accepts the context sensitive language L={$a^n b^n c^n$ n≥1} where Q={$q_0,q_1,q_2,q_3,q_{acc}$}, $Q_{acc}$={$q_{acc}$}, $Q_{rej}$={}, Σ ={a,b,c}, ρ is the identity relation. We define the linear operator in M as follows.

$V_{\#,\#}|q_0\rangle = |q_0\rangle$, $V_{\#,a}|q_0\rangle = |q_0\rangle$, $V_{\#,b}|q_0\rangle = |q_1\rangle$, $V_{a,b}|q_1\rangle = |q_1\rangle$, $V_{a,c}|q_1\rangle = |q_2\rangle$, $V_{b,c}|q_2\rangle = |q_2\rangle$, $V_{b,\$}|q_2\rangle = |q_3\rangle$, $V_{c,\$}|q_3\rangle = |q_3\rangle$, $V_{\$,\$}|q_3\rangle = |q_{acc}\rangle$, $D(q_0)=(0,1)$, $D(q_1)=(1,1)$, $D(q_2)=(1,1)$, $D(q_3)=(1,0)$, $D(q_{acc})=(0,0)$.

By inspection we see that $V_{\sigma,\tau}$ is well-formed. The automaton checks the number of a's in the first tape with the number of b's in the second tape again repeats the procedure for number of b's and c's. The above automaton accepts a string in the language with probability 1 and also rejects a string not in the language with probability 1.

## 4. MATRICES REPRESENTATION OF DIFFERENT AUTOMATON

In [21], they represent different automata using matrices and describe their properties. In this section we take transition matrices of some automaton from paper [21] and discuss different properties of these automata in terms of their transition matrices.

### 4.1 1-way multihead deterministic finite state automaton

A 1-way k-head deterministic finite state automaton is a deterministic finite state automaton with k- independent reading heads on a single input tape with end markers. On each move the machine can simultaneously read the k input cells scanned by k-heads, move each head one square to the right or keep stationary.

**Definition 4**. *A 1-way multihead deterministic finite state automaton (1DFA(k))[14] is a tuple $M = (Q, \Sigma, k, \delta, \#, \$, q_0, F)$ where*

a. *Q is a finite set of states,*

b. *Σ is an input alphabet,*

c. *$k \geq 1$ is the number of heads.*

d. *$\delta: Q \times \Sigma \cup \{\#,\$\})^k \to Q \times \{0,1\}^k$ is the partial transition function;where 1 means to move the head one square to the right and 0 means to keep the head on the current square,*

e. $\# \notin \Sigma$ is the left and $\$ \notin \Sigma$ is the right endmarkers.

f. $q_o \in Q$ is a starting state,

g. $F \in Q$ is a set of final or accepting states.

We define a 1DFA(2) $M = (\{q_0, q_1, q_2\}, \{a, b\}, 2, \delta, \#, \$, q_0, \{q_2\})$ shown in **Figure 1** which accept $L = \{a^n b^n | n \geq 1\}$ where

$$\delta(q_0, \{b, a\}) = q_1$$
$$\delta(q_1, \{b, a\}) = q_1$$
$$\delta(q_1, \{a, b\}) = q_2$$
$$\delta(q_2, \{\$, b\}) = q_2$$

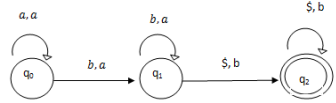

Figure 1: 1DFA(2) accept a language $L = \{a^n b^n | n \geq 1\}$

$$M_{aa} = \begin{array}{c} \\ q_0 \\ q_1 \\ q_2 \end{array} \begin{array}{c} q_0 \; q_1 \; q_2 \\ \begin{pmatrix} 1 & 0 & 0 \\ 0 & 0 & 0 \\ 0 & 0 & 0 \end{pmatrix} \end{array} \quad M_{ba} = \begin{array}{c} \\ q_0 \\ q_1 \\ q_2 \end{array} \begin{array}{c} q_0 \; q_1 \; q_2 \\ \begin{pmatrix} 0 & 1 & 0 \\ 0 & 1 & 0 \\ 0 & 0 & 0 \end{pmatrix} \end{array} \quad M_{\$b} = \begin{array}{c} \\ q_0 \\ q_1 \\ q_2 \end{array} \begin{array}{c} q_0 \; q_1 \; q_2 \\ \begin{pmatrix} 0 & 0 & 0 \\ 0 & 0 & 1 \\ 0 & 0 & 1 \end{pmatrix} \end{array}$$

Figure 2: the transition matrix of 1DFA(2) accept a language $L = \{a^n b^n | n \geq 1\}$

### 4.2 1-way Reversible multihead finite state automaton

**Definition 5.** *A 1-way reversible multihead finite state automaton (1REV-DFA(k)) [14] is a 8 tuple $M = (Q, \Sigma, k, \delta, \#, \$, q_0, F)$ which has same structure as 1DFA(k) where*

a. $Q$ is a finite set of states,

b. $\Sigma$ is an input alphabet,

c. $k \geq 1$ is the number of heads.

d. $\delta: Q \times (\Sigma \cup \{\#, \$\})^k \to Q \times \{0,1\}^k$ is the partial transition function; where 1 means to move the head one square to the right and 0 means to keep the head on the current square,

e. $\# \notin \Sigma$ is the left and $\$ \notin \Sigma$ is the right endmarkers.

f. $q_o \in Q$ is a starting state,

g. $F \in Q$ is a set of final or accepting states.

Let M be a 1DFA(k) and D be the set of all reachable configuration that occur in any computation of M beginning with an initial configuration and $(w, q, (p_1, \ldots, p_k)) \in D$ with $w = \sigma_1, \ldots, \sigma_n, \sigma_0 = \#$ and $\sigma_{n+1} = \$$. The set of all reachable configurations is denoted by D that occurs in any computation. M is said to be reversible if the following two conditions are fulfilled:

1. For any two transitions:

$$\delta(q, (\sigma_1, \ldots, \sigma_n)) = (q_1, (d_1, \ldots, d_k))$$

and

$$\delta(q', (\sigma_1, \ldots, \sigma_n)) = (q_1, (d_1', \ldots, d_k'))$$

it holds if $(d_1, \ldots, d_k) = (d_1', \ldots, d_k')$.

2. There is at most one transition of the form

$$\delta\left(q',\left(x_{p_1-d_1},\ldots,x_{p_k-d_k}\right)\right) = (q,(d_1,\ldots,d_k)) \ .$$

The non-context free language L= {$a^n b^n c^n$, n≥1} is accepted by REV-1DFA(2)
M = ($\{q_0, q_1, q_2, q_3, q_f\}, \{a, b, c\}, 2, \delta, \#, \$, q_0, \{q_f\}$) shown in **Figure 3** where the transition function $\delta$ is as follows:

$$\delta(q_0, \#, \#) = (q_0, 0, 1)$$
$$\delta(q_0, \#, a) = (q_0, 0, 1)$$
$$\delta(q_0, \#, b) = (q_1, 1, 1)$$
$$\delta(q_1, a, b) = (q_1, 1, 1)$$
$$\delta(q_1, a, c) = (q_2, 1, 1)$$
$$\delta(q_2, b, c) = (q_2, 1, 1)$$
$$\delta(q_2, b, \$) = (q_3, 1, 0)$$
$$\delta(q_3, c, \$) = (q_3, 1, 0)$$
$$\delta(q_3, \$, \$) = (q_f, 0, 0)$$

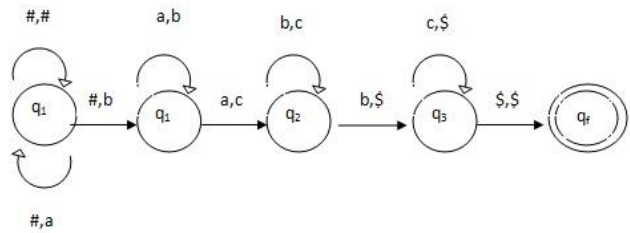

Figure 3: 1REV-DFA(2) accept a language L={$a^n b^n c^n$, n≥1}

The transition matrix of the above automaton is shown in **Figure 4**.

Figure 4: The transition matrix of 1REV-DFA(2) accept a language L={$a^n b^n c^n$, n≥1}

Dot product of any two rows is zero for multihead reversible finite state automaton.

### 4.3 2-tape 1-way quantum finite state automaton

The context sensitive language L={$a^n b^n c^n$, n≥1} is accepted by 2-tape 1-way quantum finite state automaton M=({$q_0,q_1,q_2,q_3,q_{acc}$},{$q_{acc}$}, {}, { $q_0$},{a,b,c}, $\delta,\rho$). Here $\rho$ is the injective relation between the alphabets of the input tapes (see Figure 7) where

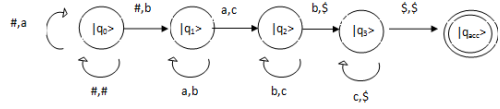

Figure 5: 2T1QFA (2) accept a language $\{a^n b^n c^n, n \geq 1\}$ with acceptance probability p>0

$$\delta(|q_0>, \#, \#) = (|q_0>, 0, 1)$$
$$\delta(|q_0>, \#, a) = (|q_0>, 0, 1)$$
$$\delta(|q_0>, \#, b) = (|q_1>, 1, 1)$$
$$\delta(|q_1>, a, b) = (|q_1>, 1, 1)$$
$$\delta(|q_1>, a, c) = (|q_2>, 1, 1)$$
$$\delta(|q_2>, b, c) = (|q_2>, 1, 1)$$
$$\delta(|q_2>, b, \$) = (|q_3>, 1, 0)$$
$$\delta(|q_3>, c, \$) = (|q_3>, 1, 0)$$
$$\delta(|q_3>, \$, \$) = (|q_{acc}>, 0, 0)$$

The transition matrix of the above automaton is shown in figure 6.

Figure 8: The transition matrix of 2T1QFA(2) accept a language $\{a^n b^n c^n, n \geq 1\}$

The sum of the square of the norms in each row adds up to 1 and dot product of any two rows is zero for 2-tape 1-way quantum finite state automaton. So if no superposition of states are involved in 2T1QFA(2), then the transition matrix of 2T1QFA(2) is similar to the transition matrix of 1RMFA(2).

## 5. COMPUTATIONAL COMPLEXITY OF 2-TAPE 1-WAY QUANTUM FINITE STATE AUTOMATON

**Theorem 1:** 1QFA(2) can accept all unary regular languages.
**Proof.** The proof is in [21].

**Theorem 2:** For every deterministic finite state automaton $D = (Q, \Sigma, q_0, F, \delta)$, we can obtain a 2-tape 1-way quantum finite state automaton $M = (Q', \Sigma', \delta', q_0', Q_{acc}, Q_{rej}, \rho)$ such that L(D)=L(M) where L(D) is the language accepted by the automaton D and L(M) is the language accepted by the automaton M.

**Proof:** To prove the theorem, in the first part we give a construction to obtain 2-tape 1-way quantum finite state automaton M from deterministic finite state automaton D. In the second part, we show that language accepted by M are same as language accepted by D.

**First Part**: The input alphabets of the two automatons are the same The other components of M are constructed as follows:
- The start state of M is $q_0'$, the set of states of M is $Q'=Q\cup\{q_0', q_{acc}\}$ where the set of accepting states i.e. $Q_{acc}=\{q_{acc}\}$ and the set of rejecting states of M is $Q_{rej}=\{\}$.
- For every $p\in \Sigma$, we form a list containing all transitions of D involving p. We arranged these transitions in any particular order and assign a number of the form $p_i$ based on its position in the list.
- If there are n transitions in the list, then we introduce the symbols $p,p_1,......,p_n$ in $\Sigma'$ and the relations $(p,p_1),(p,p_2),...,(p,p_n)$ in $\rho$.
- For a transition $\delta(q,p)=q'$ in D having number $p_i$ associated with it, we introduce the transition $V_{p,p_i}|q\rangle=|q'\rangle$ in $\delta'$. This is repeated for every transition in the list. Initially both heads of M are in # and at the end both heads of M are in \$. So we introduce two transitions which are also added to $\delta'$.'

$$V_{\#,\#}|q_0'\rangle=|q_0\rangle$$
$$V_{\$,\$}|q\rangle=|q_{acc}\rangle \text{ for all } q\in F.$$

- The both head will move one square to the right for all states other than final states i.e. D(q)=(1,1) for all $q\in Q'-Q_{acc}-Q_{rej}$. The heads are stationary if the automaton reaches the accepting states i.e. $D(q_{acc})=(0,0)$.
- As we stated in the definition of 2T1QFA(2) that the set of transitions which are not defined in $\delta'$ goes to some rejecting states.

**Second Part:** In this part we will show that M accepts the same language as D. Suppose D accepts $w$. We design relation $\rho$ in such a way that the string $w'$ of M guesses the transitions that D takes to accept $w$. Each position of $w'$ guesses the transition that D takes on reading that particular position in $w$. M simulates the transition sequence of D based on the sequence in $w'$. Since string $w$ is accepted by the automaton D, we get a sequence of transitions that takes D to the final state after consuming $w$. So, the automaton M rightly guess that particular sequence of transitions from one of the many possible $w'$ that will take the automaton D to its final state after consuming $w$. Since the automaton M simulates the automaton D, M will reach the final state of D based on that particular string of $w$. Now both heads of the automation M will reach position '\$' which will take the automation M to its accepting state $q_{acc}$. Thus the string $w$ is accepted by the automaton M by using the transition $U_{\$,\$}|q\rangle=|q_{acc}\rangle$ for all $q\in F$.

Now, we consider a string $w$, which is not accepted by the automaton D. So for such a string $w$ which is not accepted by D, we will not get such a sequence of transitions for which the automaton will reach to final state after consumption of $w$. Since the automaton M simulating the automaton D based on the string of $w'$, will never reach the situation where both heads of the automaton M are in '\$' and transitions of the form $U_{\$,\$}|q\rangle=|q_{acc}\rangle$ for all $q\in F$ cannot be applied to M. So, the automaton M will never reach the final state $q_{acc}$. As we know from the definition of M that we consider 1-way automaton i.e., the heads of the automaton move only in the direction right to the input tape, so we will reach in such a situation for which no transition rule is defined then the automaton M will go to the rejecting state. Thus, M will reject $w$.

**Corollary 1:** 2-tape 1-way quantum finite state automata with non-injective relation can accept all regular languages.

*Example 1:* A 2-tape 1-way quantum finite state automaton $M = (Q, Q_{acc}, Q_{rej}, |q_0>, \Sigma, \delta, \rho)$ can accept the language $L = \{\%w_1*x_1\%w_2*x_1...\%w_n*x_n/n\geq 0, w_i \in\{a,b\}^*, x_i\in \{a,b\}^*, \exists i \exists j : w_i=w_j, x_i\neq x_j\}$ in the following manner:
Let, $Q=\{q_0, q_1, q_2, q_3, q_4, q_5\}$, $Q_{acc}=\{q_5\}$, $Q_{rej}=\{q_4\}$, $\Sigma =\{a,b,v_{p1},v_{p2},\%,*\}$ $\rho=\{(a,a),(\%,\%),(\%,v_{p1}),(\%,v_{p2}),(b,b),(*,*)\}$.

*We define the transitions of M as follows:*

$V_{\#,\#}|q_0\rangle=|q_0\rangle$, $V_{\%,\%}|q_0\rangle=|q_0\rangle$, $V_{a,a}|q_0\rangle=|q_0\rangle$, $V_{b,b}|q_0\rangle=|q_0\rangle$, $V_{*,*}|q_0\rangle=|q_0\rangle$, $V_{\%,v_{p1}}|q_0\rangle=|q_1\rangle$, $V_{\%,a}|q_1\rangle=|q_1\rangle$, $V_{\%,b}|q_1\rangle=|q_1\rangle$, $V_{\%,*}|q_1\rangle=|q_1\rangle$, $V_{\%,\%}|q_1\rangle=|q_1\rangle$, $V_{\%,v_{p2}}|q_1\rangle=|q_2\rangle$, $V_{a,a}|q_2\rangle=|q_2\rangle$, $V_{b,b}|q_2\rangle=|q_2\rangle$, $V_{*,*}|q_2\rangle=|q_3\rangle$, $V_{b,b}|q_3\rangle=|q_3\rangle$, $V_{a,a}|q_3\rangle=|q_3\rangle$, $V_{\%,\%}|q_3\rangle=|q_4\rangle$, $V_{\%,\$}|q_3\rangle=|q_4\rangle$, $V_{a,b}|q_3\rangle=|q_5\rangle$, $V_{a,*}|q_3\rangle=|q_5\rangle$, $V_{a,\%}|q_3\rangle=|q_5\rangle$, $V_{a,\$}|q_3\rangle=|q_5\rangle$, $V_{b,a}|q_3\rangle=|q_5\rangle$, $V_{b,*}|q_3\rangle=|q_5\rangle$, $V_{b,\%}|q_3\rangle=|q_5\rangle$, $V_{b,\$}|q_3\rangle=|q_5\rangle$, $V_{*,a}|q_3\rangle=|q_5\rangle$, $V_{*,b}|q_3\rangle=|q_5\rangle$, $V_{*,\%}|q_3\rangle=|q_5\rangle$, $V_{*,\$}|q_3\rangle=|q_5\rangle$, $V_{\%,a}|q_3\rangle=|q_5\rangle$, $V_{\%,b}|q_3\rangle=|q_5\rangle$, $V_{\%,*}|q_3\rangle=|q_5\rangle$.

$D(q_0)=(1,1)$, $D(q_1)=(0,1)$, $D(q_2)=(1,1)$, $D(q_3)=(1,1)$, $D(q_4)=(0,0)$, $D(q_5)=(0,0)$.

*The above stated automaton works in the following manner:*

*Initially both heads of the automaton M are at #. Since the start state of the automaton is $q_0$, after reading the inputs symbols, the automaton M remains in state $q_0$ but both heads of the automaton will move one square to the right. In this example, we use $(\%,v_{p1})$ and $(\%, v_{p2})$ elements to guess the two substrings of the input string which has its w parts equal and x parts*

*unequal. The automaton goes to state $q_2$ after finding the guessed substring. In state $q_2$, if the substring do not have their w parts equal then the automaton goes to the rejecting state as no transition rules are defined for such a situation in state $q_2$ and the automaton rejects the input string with probability 1. Now if the two guessed substrings have w parts equal then the automaton goes to state $q_3$. In state $q_3$, the automaton M checks if w parts of the guessed substrings are equal then whether x parts are equal or not. The automaton M goes to state $q_4$ if x parts of the guessed substrings is also equal. Since $q_4$ is a rejecting state of the automaton M thus the automaton rejects the input string with probability 1. Now the automaton goes to state $q_5$ if x parts are unequal. $q_5$ is the accepting state of the automaton M. Thus, the input string is accepted with probability 1.*

*We consider a string p which is not in L. One of the many strings obtained by applying ρ on p will correctly guess the two substrings which have their w parts equal and x parts unequal and the automaton M will accepts the string p with probability 1.*

*Now consider a string p which is not in L. Since p is not in L, so there are no two substrings which have their equal w parts and unequal x parts. Therefore, there are no strings obtained by applying ρ on p which will correctly guess the location of the two substrings of p for which their w parts are equal and x parts are unequal. So the automaton M rejects p with probability 1. Thus from the above stated arguments we can conclude that the automaton M accepts L.*

**Lemma 1:** The language L = {%$w_1$*$x_1$%$w_2$*$x_1$...%$w_n$*$x_n$|n≥0, $w_i$ ∈{a,b}$^*$, $x_i$∈ {a,b}$^*$, ∃i∃j :$w_i$=$w_j$, $x_i$≠$x_j$} is not accepted by any deterministic multi-head finite automaton.

The proof of Lemma 1 is in Yao et. al.[22]

**Theorem 3:** $L_{2T1QFA}$-$L_{DFA(k)}$≠ ∅, where $L_{2T1QFA}$ is the set of all languages accepted by 2-tape 1-way quantum finite automata and $L_{DFA(k)}$ is the set of all languages accepted by multi-head deterministic finite automata.

**Proof:** From Example 1, we know that there is a 2-tape 1-way quantum finite automaton that accepts the language L = {%$w_1$*$x_1$%$w_2$*$x_1$...%$w_n$*$x_n$|n≥0, $w_i$ ∈{a,b}$^*$, $x_i$∈ {a,b}$^*$, ∃i∃j :$w_i$=$w_j$, $x_i$≠$x_j$} and from Lemma 1 we know that L = {%$w_1$*$x_1$%$w_2$*$x_1$...%$w_n$*$x_n$|n≥0, $w_i$ ∈{a,b}$^*$, $x_i$∈ {a,b}$^*$, ∃i∃j :$w_i$=$w_j$, $x_i$≠$x_j$} is not accepted by any deterministic multi-head finite automaton which proves the above Theorem.

**Corollary 2:** The set of languages accepted by one-way reversible multi-head finite automata with two heads is a proper subset of set of languages accepted 2-tape 1-way quantum finite state automata.

**Proof:** We know that the transition matrix of 1-way reversible multihead finite state automaton has the following properties:
(1)Dot product of any two row is zero for 1-way reversible multihead finite state automaton.
(2)All matrices only have 0 or 1 entries.
Therefore the above two properties of the transition matrix ensures that the transition matrix is also unitary. As a result given a 1-way reversible 2-head finite state automaton M we get a 2-tape 1-way quantum finite state automaton M' which has the same transition matrix, same set of states, same set of accepting states and start state as M. As the transition matrix, start state and accepting states of M and M' are same, they accept the same language. So, for every 1RMFA(2) which accept a language L there exist 1QFA(2) which accept the same language. So,the set of all languages accepted by 1RMFA(2) is a subset of set of all languages accepted by 1QFA(2)

Moreover, it has already been stated by Kutrib et.al.[13] that set of languages accepted by multi-head reversible finite automata is a proper subset of set of languages accepted by multi-head deterministic finite automata. Thus there is no multi-head reversible finite automaton which accept the language L = {%$w_1$*$x_1$%$w_2$*$x_1$...%$w_n$*$x_n$|n≥0, $w_i$ ∈{a,b}$^*$, $x_i$∈ {a,b}$^*$, ∃i∃j :$w_i$=$w_j$, $x_i$≠$x_j$} but from Theorem 5, we see that a 2-tape 1-way quantum finite state automaton can accept the language L. Thus, the subset relation is proper which proves the Corollary.

**Theorem 4:** There is a 2-tape 1-way quantum finite state automaton that accepts the context sensitive language L={ww |w∈ {a, b}$^*$}.

**Proof:** M=(Q, V, δ, $q_0$, $Q_{acc}$, $Q_{rej}$, ρ) is a 2-tape 1-way quantum finite state automaton with non-injective relation ρ that accepts the context sensitive language L={ww |w∈ {a, b}$^*$} where Q={$q_0$, $q_1$, $q_2$, $q_3$, $q_4$, $q_5$, $q_6$, $q_7$, $q_8$, $q_{rej}$, $q_{rej1}$, $q_{rej2}$, $s_1$, $s_2$}, $Q_{acc}$={$s_2$}, $Q_{rej}$={$s_1$, $q_{rej}$, $q_{rej1}$, $q_{rej2}$}, Σ ={a, b, m}, ρ={(a,a), (a,m), (b,b), (b,m)}. We define the transitions involved in M as follows:

$V_{\#,\#}|q_0\rangle=|q_0\rangle$, $V_{\#,a}|q_0\rangle=|q_0\rangle$, $V_{\#,b}|q_0\rangle=|q_0\rangle$, $V_{\#,m}|q_0\rangle=\frac{1}{\sqrt{2}}|q_1\rangle+\frac{1}{\sqrt{2}}|q_2\rangle$, $V_{\#,m}|q_1\rangle=|q_3\rangle$, $V_{a,a}|q_3\rangle=|q_3\rangle$, $V_{b,b}|q_3\rangle=|q_3\rangle$, $V_{a,b}|q_3\rangle=|q_{rej}\rangle$, $V_{b,a}|q_3\rangle=|q_{rej}\rangle$, $V_{x,\$}|q_3\rangle=|q_4\rangle$ x∈{a,b}, $V_{x,\$}|q_4\rangle=|q_4\rangle$ x∈{a,b}, $V_{\$,\$}|q_4\rangle=|q_5\rangle$, $V_{\#,m}|q_2\rangle=|q_6\rangle$, $V_{x,y}|q_6\rangle=|q_7\rangle$ x,y∈{a,b}, $V_{x,y}|q_7\rangle=|q_6\rangle$ x,y∈{a,b}, $V_{x,\$}|q_6\rangle=|q_{rej2}\rangle$ x∈{a,b}, $V_{\$,x}|q_6\rangle=|q_{rej2}\rangle$ x∈{a,b}, $V_{\$,x}|q_7\rangle=|q_{rej1}\rangle$ x∈{a,b}, $V_{x,\$}|q_7\rangle=|q_{rej1}\rangle$ x∈{a,b}, $V_{\$,\$}|q_7\rangle=|q_8\rangle$, $V_{\$,\$}|q_5\rangle=\frac{1}{\sqrt{2}}\sum_{l=1}^{2}e^{\frac{2\pi i}{2}\cdot 1 \cdot l}|s_l\rangle$, $V_{\$,\$}|q_8\rangle=\frac{1}{\sqrt{2}}\sum_{l=1}^{2}e^{\frac{2\pi i}{2}\cdot 2 \cdot l}|s_l\rangle$. D($q_0$)=(0,1), D($q_1$)=(0,0), D($q_2$)=(0,0), D($q_3$)=(1,1), D($q_4$)=(1,0), D($q_5$)=(0,0), D($q_6$)=(1,1), D($q_7$)=(1,0), D($q_8$)=(0,0), D($s_1$)=(0,0), D($s_2$)=(0,0), D($q_{rej}$)=(0,0), D($q_{rej1}$)=(0,0), D($q_{rej2}$)=(0,0).

The above mentioned automaton works in the following manner:

The automaton M works in three phases. In the first phase, the elements (a, m) or (b, m) of the relation ρ defined on input alphabet is used to guess the end of first word w in the second tape. On finding m, the automaton M goes to the second phase. In the second phase the computation branches into 2 paths, indicated by the states $q_1$ and $q_2$ each with amplitude $\frac{1}{\sqrt{2}}$. In each of these two paths, the two tape heads of each individual path move deterministically from the current position to the right end marker '$'  independently. The first path checks whether the string in the first tape after # to the position of m is same as the string in the second tape after m to $. If some character is not the same then this path ends in rejecting state (i.e. the path verifies whether input is of the form ww ). At the same time, in the second path every time first tape head is moved two steps, the second tape head is moved one step, to check whether position of 'm' at the end of first word w has been correctly guessed. Only if the position of 'm' is correctly guessed will the two heads of the second path reach '$' at the same time otherwise only one head of the second path goes to '$' and the computation in the second path halts in a rejecting state.

In the third phase, when both the heads of the individual paths arrive at '$' computation in each path again splits according to the quantum Fourier transform yielding either the single accepting state $s_2$ with probability 1 or a rejecting state with probability at least $\frac{1}{2}$.

Now consider a string s in L. As s is in L, s is of the form ww, one of the many strings of s will guess the position of m at the end of first w correctly, let that string be s'. All the four heads of the two individual paths of automaton M with s in the first tape and s' in the second tape will reach '$' at the same time, by the superposition of the machine immediately after performing quantum Fourier transform we get $\frac{1}{2}\sum_{j=1}^{2}\sum_{l=1}^{2}(e^{\frac{2\pi i}{2}\cdot j\cdot l})|s_l\rangle=|s_2\rangle$. Hence the observable yields the result, accept with probability 1

For a string s not in L, s is not of the form ww, so no matter the position guessed by any string of w it can never be at the end of first w. As a result, at least one of the 4 heads of two individual paths of M, will not reach $ and the superposition will not result in the accepting state $s_2$. There will be a presence of a rejecting state with probability of atleast $\frac{1}{2}$. Thus M reject w with a probability of at least $\frac{1}{2}$.

## 6. CONCLUSION

We have shown that in spite of 2-tape 1-way quantum finite automata being reversible in nature they accept all regular languages. We have also explored and compared the computational power 2-tape 1-way quantum finite automata with other existing deterministic and reversible automata models and utilized the superposition principle to show acceptance of the language L={ww |w∈ {a, b}$^*$} by 2-tape 1-way quantum finite state automata. We have also established that our Quantum model accept languages which are not accepted by any multi-head deterministic finite automata.

### Acknowledgments
Research of Debayan Ganguly is funded by the Council of Scientific Industrial Research (CSIR). This support is greatly appreciated.